\documentclass[aps,prl,twocolumn,groupedaddress,10pt]{revtex4-1}

\usepackage{amsmath,amssymb,bm,graphicx,color,comment}

\begin{document}

\title{Universal hypotrochoidic law for random matrices with cyclic correlations}

\author{Pau Vilimelis Aceituno}
\email[]{pau.aceituno@mis.mpg.de}
\affiliation{Max Planck Institute for Mathematics in the Sciences, 04103 Leipzig, Germany}
\author{Tim Rogers}
\email[]{t.c.rogers@bath.ac.uk}
\affiliation{Centre for Networks and Collective Behaviour, Department of Mathematical Sciences, University of Bath, Bath, BA27AY, UK}
\author{Henning Schomerus}
\email[]{h.schomerus@lancaster.ac.uk}
\affiliation{Department of Physics, Lancaster University, Lancaster, LA1 4YB, UK}

\begin{abstract}
The celebrated elliptic law describes the distribution of eigenvalues of random matrices with correlations between off-diagonal pairs of elements, having applications to a wide range of physical and biological systems. Here, we investigate the generalization of this law to random matrices exhibiting higher-order cyclic correlations between $k$-tuples of matrix entries. We show that the eigenvalue spectrum in this ensemble is bounded by a hypotrochoid curve with $k$-fold rotational symmetry. This hypotrochoid law applies to full matrices as well as sparse ones, and thereby holds with remarkable universality. We further extend our analysis to matrices and graphs with competing cycle motifs, which are described more generally by polytrochoid spectral boundaries.
\end{abstract}


\maketitle

Determining the eigenvalue spectra of large random matrices is a rich theoretical problem \cite{anderson_guionnet_zeitouni_2009} with many applications in fields as diverse as telecommunications \cite{tulino2004random}, quantum physics \cite{guhr1998random}, ecology \cite{may1972will,allesina2012stability} and economics \cite{rosenow2000random}. A key result in this field is the elliptic law \cite{girko1986elliptic}, which states that in the limit of large matrix size the eigenvalues for random matrices with correlations between symmetric pairs of entries are confined within an ellipse in the complex plane.  This result originating from over 30 years ago still drives scientific developments today, in both mathematical theory \cite{tao2010random,naumov2012elliptic} as well as applications \cite{aceituno2017tailoring}.

As the applications of random matrix theory have diversified, so too have the ensembles under study. Existing generalizations of the elliptic law fall broadly into three categories. There is a large body of theoretical work concerning random matrix ensembles defined by potential functions, where the quadratic case recovers the elliptic law, but more complicated potentials produce interesting spectral distributions (see e.g. \cite{elbau2005density,bleher2012orthogonal}). Alternatively, for some applications it is necessary to impose a system-level structure such as modularity (see e.g. \cite{grilli2016modularity}), often by addition or multiplication with another matrix; see \cite{rogers2010universal,bordenave2011spectrum} for methods and rigorous mathematical results. Lastly, growing interest in the study of complex networks has led many authors to consider sparse ensembles containing many zero matrix elements, where the location of the non-zero elements encodes the adjacency matrix of some directed graph (or \emph{digraph}), and can have striking eigenvalue distributions \cite{metz2011spectra,bolle2013spectra} (see \cite{metz2018spectra} for a recent topical review). In none of these directions of work has the problem of high-order correlations been fully and directly addressed. This is not for lack of interest. Multi-party interactions in dense systems are important in biological applications such as ecology \cite{levine2017beyond}, stabilization of microbial communities \cite{guo2016contribution} or gene--gene interactions \cite{wang2014gene} and can provide valuable engineering insights into machine learning \cite{sejnowski1986higher,personnaz1987high} and control theory \cite{doyle1981multivariable}. Moreover, most of the sparse random matrix literature hinges on the assumption of a \emph{tree-like} interaction structure; it remains a long-standing and important problem to allow the relaxation of this assumption, inducing higher-order correlations.

\begin{figure}
\includegraphics[width=\columnwidth, trim=40 0 50 0, clip=true]{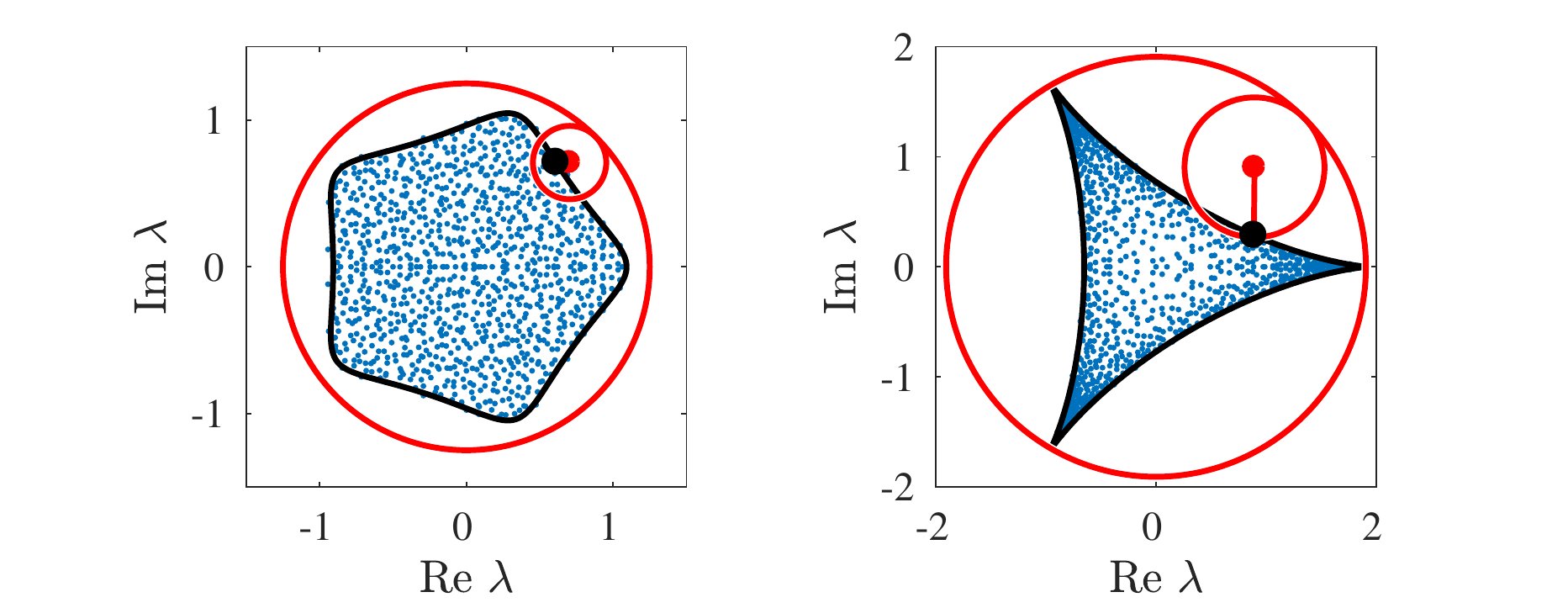}
\caption{Hypotrochoid curves (black lines) bounding the eigenvalue spectra (blue dots) of random matrices with higher-order correlations. Left: a dense $N\times N$ random matrix $M$ with $\overline{\textrm{Tr}M^5}/N=0.075$ and other correlations negligible. Right: a random digraph in which each node appears in exactly two directed cycles of length three, with no other edges. In both cases $N=1000$. The large and small red circles respectively show the fixed and rotating wheels describing the hypotrochoid curve.}
\label{hyp5}
\end{figure}

Here, we investigate the generalization of the elliptic law to ensembles, both dense and sparse, featuring higher-order correlations. In the case that there is a single dominant correlation order $k$, we find that the spectrum is bounded by a hypotrochoid curve (that is, the path of a point located on a small wheel that rolls inside a larger wheel) with $k$-fold rotational symmetry; see Fig.~\ref{hyp5} for an illustration. Surprisingly, we thereby recover the spectral boundary of a highly structured system, a sparse regular digraph \cite{metz2011spectra,bolle2013spectra}, but in a much more general and universal context. Extending this result further to the case where more than one correlation order is present, we find a more general family of \emph{polytrochoid} boundary curves described by multiple linked gears \footnote{See, for example, US patent 3037488A ``Rotary hydraulic motor'', George M Barrett, (1962)}.

\emph{Dense matrices}.---We study $N\times N$-dimensional nonhermitian matrices with real or complex entries independently drawn from a distribution with zero mean and bounded variance. For any of such matrix $M$, the density $\mu(z)$ of complex eigenvalues $z$ can be obtained from a Green's function of the form \cite{feinberg1997non,janik1999correlations,rogers2010universal}
\begin{equation}\label{eq:e}
G(z,z^*)=
           \begin{pmatrix}
             z\openone -M & i\lambda\openone \\
             \!\!\! i\lambda\openone & \!\!\!\! z^*\openone -M^\dagger
           \end{pmatrix}^{-1}
   \equiv
           \begin{pmatrix}
             G_{11} & G_{12} \\
             G_{21} & G_{22}
           \end{pmatrix}
         ,
\end{equation}
where $z^*$ denotes complex conjugation and $M^\dagger$ the adjoint; for a real matrix this is just the transpose.
Thereby
\begin{align}\label{mug}
\mu(z)&=\frac{1}{\pi}\frac{\partial g_{11}}{\partial z^*}, \\
g(z,z^*)&=\lim_{\lambda\to 0^+}
           \begin{pmatrix}
             \textrm{Tr}\,G_{11} & \textrm{Tr}\,G_{12} \\
             \textrm{Tr}\,G_{21} & \textrm{Tr}\,G_{22}
           \end{pmatrix}
         \equiv
           \begin{pmatrix}
             g_{11} & g_{12} \\
             g_{21} & g_{22}
           \end{pmatrix}.\nonumber
\end{align}
For a random matrix $M$, the ensemble-averaged density $\overline{\mu}(z)$ therefore follows from $\overline{g}(z,z^*)$.

This can be used to derive the elliptic law of random matrices in the large $N$ limit, which serves as useful preparation.
Let us expand
\begin{equation}
G=\mathcal{Z}^{-1}\sum_{\ell=0}^\infty (\mathcal{M}\mathcal{Z}^{-1})^\ell
\end{equation}
into a geometric series, where
\begin{align}
\mathcal{Z}&=Z\otimes\openone,
\quad
Z=
           \begin{pmatrix}
             z & i\lambda \\
             i\lambda & z^* \\
           \end{pmatrix},
         \quad
        % \\
         \mathcal{M}=
           \begin{pmatrix}
             M & 0 \\
             0 & M^\dagger \\
           \end{pmatrix}. %\nonumber
\end{align}
When we perform the averaging, only certain products of matrix elements $M_{nm}$ survive. These can be organised into groups of terms whose number scales differently in the matrix dimension $N$. In particular, there will be many terms where we can pair $M_{nm}$ with $(M^\dagger)_{mn}$. If no further correlations are present the leading order is
\begin{align}
\overline{G}&=\mathcal{Z}^{-1}+\mathcal{Z}^{-1}\dot{ \mathcal{M}} \overline{\sum_{\ell=0}^\infty (M\mathcal{Z}^{-1})^\ell}\dot{\mathcal{M}} \overline{\sum_{\ell=0}^\infty (M\mathcal{Z}^{-1})^\ell}\nonumber
\\
&=\mathcal{Z}^{-1}+\mathcal{Z}^{-1}\dot{ \mathcal{M}} \overline{G} \dot{\mathcal{M}} \overline{G}
,
\end{align}
where the dot denotes matrices in which we pair the elements  $M_{nm}$ with $(M^\dagger)_{mn}$. This factorization of the average is called the non-crossing or planar approximation.

The elliptic law is derived straightforwardly from the expression above. Carrying out the average of the dotted matrices, taking the partial trace on both sides, and rearranging for $Z$, one obtains
\begin{equation}
Z=N/\overline{g}+
  \left(
           \begin{array}{cc}
             \tau_2^2\overline{g}_{11} & \sigma^2 \overline{g}_{12} \\
             \sigma^2 \overline{g}_{21} & \tau_2^2\overline{g}_{22} \\
           \end{array}\right).
           \label{elli}
\end{equation}
Comparing terms in the off-diagonal in this equation (and noting the constraints $\bar g_{11}=\bar g^*_{22}$ and $g_{12}=g_{21}>0$), we find two possible solutions: (i) either $g_{12}=0$, or (ii) $|\bar g_{11}|^2-\bar g_{12}^2=N/\sigma^2$. The first of these yields $\bar g_{11}$ proportional to $z$ and therefore holds only outside of the support of the spectrum. Examining the diagonal elements of (\ref{elli}) in the case (ii) we obtain
\begin{equation}
z=\sigma^2\overline{g}_{11}^*+\tau_2^2\overline{g}_{11}\,.
\label{elli2}
\end{equation}
which must be solved together with the constraint that $|\bar g_{11}|^2-N/\sigma^2>0$. The boundary of the spectrum is therefore found by determining the values of $z$ for which $|\bar g_{11}|^2=N/\sigma^2$; in the present case one finds an ellipse with foci $\pm2\tau\sqrt{N}$. Inside the support, one can solve (\ref{elli2}) to determine $\overline{g}_{11}$ and apply (\ref{mug}) to find that the spectral density $\mu(z)$ is uniform.

To generalize these results to ensembles with high-order correlations where $\overline{\textrm{Tr}M^k}/N$ is a fixed parameter, we reinterpret the Hermitian contributions of weight $\tau_2$ in the elliptic law as correlations of order $k=2$. Introducing correlations of general order $k$, we pick up additional contributions corresponding to contractions
\begin{align}
\label{eq:gen}
\overline{G}&=\mathcal{Z}^{-1}+
\sum_{k} \mathcal{Z}^{-1}(\dot{ \mathcal{M}} \overline{G})^{k-1} \dot{\mathcal{M}} \overline{G}
.
\end{align}
We assume for now that there is just a single extra term of these, of fixed $k$ and with weight $\tau_k$.
This gives
\begin{equation}
Z=N/\overline{g}+
  \left(
           \begin{array}{cc}
             \tau_k^k\overline{g}_{11}^{k-1} & \sigma^2 \overline{g}_{12} \\
             \sigma^2 \overline{g}_{21} & \tau_k^k\overline{g}_{22}^{k-1} \\
           \end{array}\right)
\end{equation}
and results, inside the spectrum, in the equations
\begin{equation}
\begin{split}
\label{eq:cond}
z&=\sigma^2\overline{g}_{11}^*+\tau_k^k\,\overline{g}_{11}^{k-1},\\
\overline{g}_{12}^2&=|\overline{g}_{11}|^2-N/\sigma^2.
\end{split}
\end{equation}
\begin{figure}
\includegraphics[width=0.85\columnwidth, trim=0 0 0 0, clip=true]{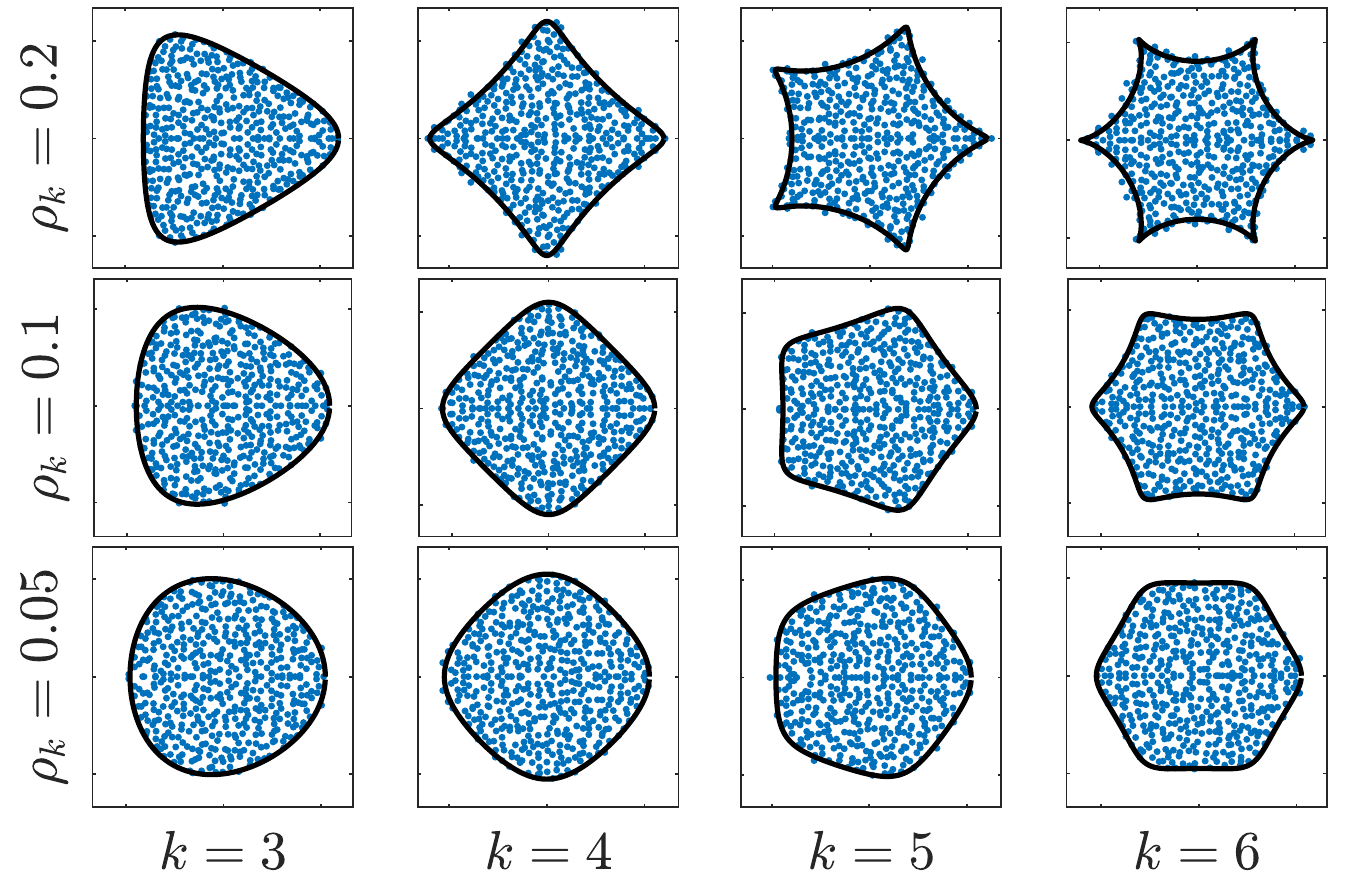}
\caption{Hypotrochoid curves (black lines) bounding the eigenvalue spectra (blue dots) of random matrices with correlations $\overline{\textrm{Tr}M^k}=N\rho_k$.}
\label{hypd}
\end{figure}
The boundary of the support is therefore determined by the condition $|\overline{g}_{11}|^2=N/\sigma^2$. In the large $N$ limit we choose the scalings $\sigma^2=N^{-1}$, $\tau_k^k=\rho_k N^{1-k}$ to balance the contribution of terms in (\ref{eq:cond}). With the parameterization $\overline{g}_{11}=Ne^{i\varphi}$,  which we insert into Eq.~\eqref{eq:cond}, the boundary curve then becomes
\begin{equation}
\label{eq:zb0}
z_{b}(\varphi)=e^{-i\varphi}+\rho_k e^{i(k-1)\varphi}.
\end{equation}

This equation is precisely the complex parameterization of a hypotrochoid curve in which the small and large wheels have radii in a ratio of $1:(k-1)$. In Fig.~\ref{hypd} we illustrate this result numerically for various values of $k$ and $\rho_k$; see Appendix A for the matrix generation algorithm.

Note that (although hard to determine from Fig.~\ref{hypd}), for $k>2$ the density of eigenvalues inside the hypotrochoid support is in fact not uniform; in general solutions to (\ref{eq:cond}) do not have the property that $\partial \bar g_{11}/\partial z^*$ is constant. This distribution is, however, universal in the sense that it is determined entirely by the parameters $\sigma^2$ and $\rho_k$, and other properties of the distribution of matrix elements are unimportant. We will now explore to what extent this universality extends to sparse matrices.

\emph{Sparse digraphs}.---
Square matrices can be seen as an alternative representation of weighted digraphs, where the entry $M_{nm}$ corresponds to the weight of the edge going from node $n$ to node $m$. This implies that for large dense graphs with random weights we can obtain the eigenvalues of their adjacency matrix by the previous methods. However, it is well-known that sparsity can change the eigenvalue distribution substantially \cite{semerjian2002sparse,biroli1999single,rogers2009cavity}. Surprisingly, the hypotrochoidic law \eqref{eq:zb0} also applies to highly-structured sparse systems. Randomly generated digraphs in which each node belongs to exactly $d$ directed cycles of length $k$ were studied in \cite{metz2011spectra,bolle2013spectra}, where it was shown that the adjacency matrices of these graphs hypotrochoidic spectra in the limit of large network size. 

\begin{figure}
\includegraphics[width=0.9\columnwidth]{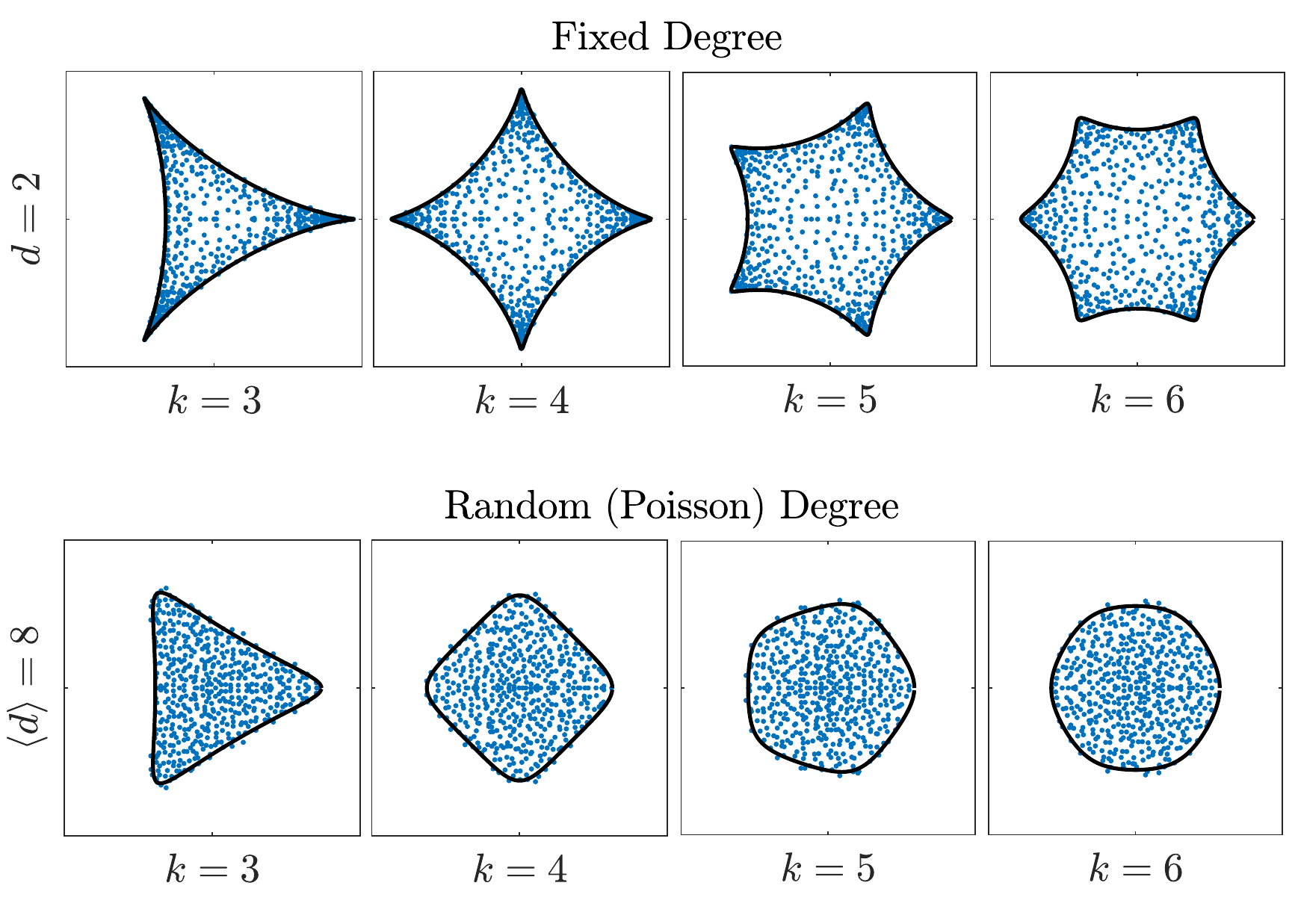}
\caption{Hypotrochoid curves (black lines) bounding the eigenvalue spectra (blue dots) of sparse random digraphs composed of $k$-cycles. In the top set of figures the networks were generated to have nodes with fixed in-degree and out-degree, here $d=2$; in the lower set nodes are assigned to cycles uniformly randomly, resulting in Poisson distributions for both in-degree and out-degree, with the mean degree being $\langle d\rangle =8$ in this case (for Poisson graphs $\hat d=\langle d\rangle$). Note the similarity to the bottom row with the plots presented in Fig.~\ref{hypd}.}
\label{hyps}
\end{figure}

This result extends further to disordered directed random graph ensembles. In Appendix B we apply effective medium approximation (EMA) \cite{dorogovtsev2003spectra} to derive the following hypotrochoidic law for the spectral boundary of cyclic random digraphs: 
\begin{equation}
z_b(\varphi)=\frac{1}{t}e^{-i\varphi}+\hat d t^{k-1}e^{i(k-1)\varphi},
\label{eq:zb2}
\end{equation}
where $k$ is the length of cycles, $\hat d$ (degree biased) number of cycles per node, and $t$ is the unique positive real solution of $\hat dt^{2k}-dt^2+1=0$. The EMA is technically valid in the case $1\ll d\ll N$, however, numerical simulations show excellent agreement down to relatively small values of $d$, see Fig.~\ref{hyps}. For fixed node degrees it is exact.

To understand how both sparse and dense cyclic ensembles share the same universal behaviour, we explore the asymptotic behaviour of (\ref{eq:zb2}) as $\hat d\to\infty$. For a direct comparison we rescale the adjacency matrix of the graph by $\hat d^{-1/2}$, which corresponds to the factor $\sigma=N^{-1/2}$ that we used before and that normalizes rows and columns. The previous result \eqref{eq:zb2}  for $k=3$ then reads
\begin{equation}
z_{b}(\varphi)=\hat d^{-1/2}t^{-1}e^{-i\varphi}-(\hat d^{1/2}-\hat d^{-1/2})t^2 e^{2i\varphi}
\end{equation}
with $t$ being the same as in Eq. \eqref{eq:zb2}. We note that for large $\hat d$, $t\sim \hat d^{-1/2}$, so that we recover the support \eqref{eq:zb0} for full matrices with an effective parameter $\rho_3\sim \hat d^{-1/2}$.

The same connection indeed appears when we compare the definition of the quantities $\sigma_2$ and $\rho_\tau$ from the traces of $M$. By applying the noncrossing approximation to the full matrix we find
\begin{align}
\lim_{N \to\infty}\frac{1}{N}
\overline{\textrm{Tr}\,M^{3l}}&= \frac{1}{2n+1}\binom{3l}{l}\rho_3^{l}, \\
\lim_{N \to\infty}\frac{1}{N}
\overline{\textrm{Tr}\,(MM^{\dagger})^{l}}&=\frac{1}{l}\binom{2l}{l},
\end{align}
while carrying out the corresponding
combinatorics for the graphs gives
\begin{align}
\lim_{N \to\infty}\frac{1}{N}
\overline{\textrm{Tr}\,M^{3l}}&= A^{(3)}(l,\hat d)\hat d^{-3l/2},\\
\lim_{N \to\infty}\frac{1}{N}
\overline{\textrm{Tr}\,(MM^{\dagger})^{l}}&= A^{(2)}(l,\hat d)\hat d^{-l},
\end{align}
which we express in terms of the number of number of self-returning walks of length $2l$ from the root of an infinite $k$-regular tree:
\begin{equation}
A^{(m)}(l,\hat d)=\frac{d}{l}\sum_{j=0}^{l-1}\binom{ml}{j}(l-j)(\hat d-1)^j.
\end{equation}
Asymptotically for large $\hat d$, $A^{(m)}(l,\hat d)= \frac{\hat d^l}{l}\binom{ml}{l-1}=\frac{\hat d^l}{ml-l+1}\binom{ml}{l}$, so that both expressions again match up for $\rho_3\sim \hat d^{-1/2}$. It is worth mentioning that the traces of $M$, which can be computed explicitly, relate to asymptotic spectral statistics in the so-called "trace formulas" which are relevant in random matrix theory \cite{haake1996secular} as well as in semiclassical physics \cite{haake2013quantum,gutzwiller2013chaos}.

\emph{Polytrochoid spectra}.---
Starting from Eq.~\eqref{eq:gen}, our approach generalizes to matrices with correlations of multiple orders, leading to a boundary curve
\begin{equation}\label{eq:fullMtxMultipleK}
z_{b}(\varphi)=e^{-i\varphi}+\sum_k\rho_ke^{i(k-1)\varphi}.
\end{equation}
The curve described by this equation is an example of the very general polytrochoid family. For the particular case of two competing correlation orders, the curve is described by the tracing the path of a point in a wheel rotating around a larger wheel, which itself is rotating \emph{in the opposite direction} around the origin. Adding further correlation orders would correspond to the addition of more linked wheels with the same drawing procedure.

\begin{figure}
\includegraphics[width=0.9\columnwidth]{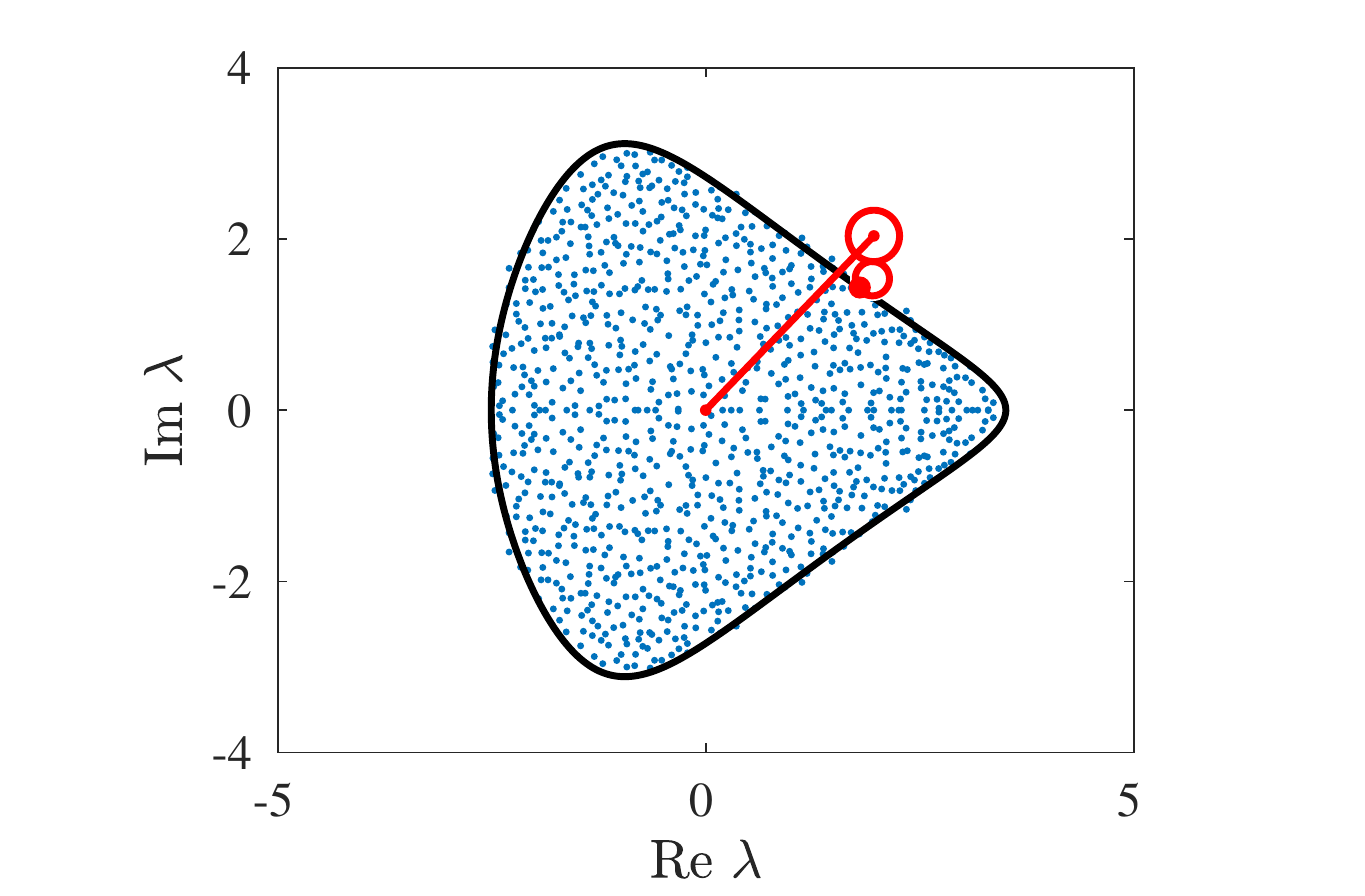}
\caption{Polytrochoid curve (black line) bounding the eigenvalue spectrum (blue dots) of a sparse regular random digraphs composed of 3-cycles and 4-cycles, with each node appearing in four of each. In this case the curve is traced by the dot marked on the smaller wheel, which makes three turns around the larger wheel while the larger wheel makes one turn around the orgin, in the opposite direction. }
\label{poly}
\end{figure}

Polytrochoid spectral boundaries are also found in digraphs with mixed cycle lengths, where each vertex connects to $d_k$ $k$-cycles of weight $w_k$; Appendix D for a derivation in the case of two competing cycle motifs. In the limit of large degrees, we obtain the explicit formula
\begin{equation}
\frac{z}{\bar d}=e^{-i\varphi}+d_1\left(\frac{w_1}{\bar d}\right)^{k_1}e^{i(k_1-1)\varphi}
+d_2\left(\frac{w_2}{\bar d}\right)^{k_2}e^{i(k_2-1)\varphi},
\end{equation}
where $\bar d=\sqrt{d_1w_1^2+d_2w_2^2}$. Fig.~\ref{poly} shows a numerical illustration of this result, which is in excellent agreement with our derivations for relatively degrees.

\emph{Discussion}.---
We have shown that high-order correlations in the entries of a random matrix give a surprising and even beautiful shape to its eigenvalues, with the boundary given by a hypotrochoid, generalizing the classical result from Girko \cite{girko1986elliptic}. Furthermore, we have uncovered a remarkable degree of universality of this result by connecting it to the known case of regular graphs with cyclic motifs\cite{metz2011spectra}, and studied the relationship between both cases. Our derivations are in excellent agreement with numerical results.

Our results have a simple interpretation in terms of systems theory. A feedback loop is a classical control theory tool to enhance or dampen certain frequencies. A cycle of length $k$ where the product of the edge weights is $w$ is a feedback loop with delay $k$ and weight $w$, therefore a graph with abundance of cycles of length $k$ with positive feedback would resonate at a frequency $\frac{1}{k}$. On the spectral side, the presence of large positive correlations of order $k$ leads to dominant eigenvalues (or poles, in control theory terms) with phases $\frac{2\pi j}{k}$ for $j<k$.

This interpretation can also be used in the other direction. Matrices and graphs are useful models to represent the interactions between elements, and our results show that random matrix theory can account for intricate multi-element interactions. That is, the stability and resonance properties of complex systems with interconnected feedback loops can be studied by converting the loops into graph cycles and then applying our results in random matrix theory.
For example, designing large networked systems such as the Internet or power grids is a challenging problem due to the amount of feedback loops present \cite{fairley2004unruly,low2002internet}. Likewise, biological regulatory systems \cite{thomas1995dynamical, becskei2000engineering,csete2004bow} include many intertwined feedback loops that render their analysis difficult. In all those examples, their stability and dynamical properties can be studied through the spectrum of their adjacency matrix, which can be difficult to estimate. However, finding frequent cycles and their corresponding delays is typically easier \cite{milo2002network,shen2002network}, meaning that we can leverage the simplicity of finding cycles to obtain rigorous results on the stability of those systems. We hope that the techniques and results presented here may inspire new developments in these fields.

Finally, the system theory interpretation also provides relevant theoretical insights and questions. For instance, the effects of negative correlations, which in systems theory corresponds to a rotation of the poles, is also explained by Eq. \eqref{eq:zb0} and \eqref{eq:zb2}. On the other hand, the combination of cycles with positive and negative feedbacks for the same length is not fully covered by our method: in the case of a full matrix or a dense digraph, the positive and negative feedbacks appear in a single connected graph and thus cancel each other -- meaning that Eq. \eqref{eq:zb0} and \eqref{eq:zb2} remain valid --, but in very sparse digraphs the presence of many finite-size isolated subgraphs implies that there are components dominated by either positive or negative feedback, and thus the effective medium approximation does not hold.

\begin{acknowledgments}
This work was supported by the Royal Society (TR) and EPSRC (HS) via grant EP/P010180/1. The authors are grateful to Izaak Neri for highlighting important prior work on sparse networks with cycles. 
\end{acknowledgments}

\appendix

\section{Appendix A: Generation of Matrices with high-order correlations}
The algorithm to generate a random matrix with positive correlations of order $k$ proceeds by going through the nodes $n$ one by one and modifying only the weights of the edges between node $n$ and nodes $<n$. In more precise terms:
\begin{enumerate}
	\item Generate a $N\times N$ matrix $M$ with random i.i.d entries. Set $n=k-1$
	\item Take the square submatrix $M_{n\times n}$ which contains the first $n$ entries of the first $n$ rows.
	\item Then compute the matrix $P_{n,k-1}$ of weights of paths of length $k-1$ between all pairs of nodes that do not pass twice by the same node.
	\item Compute the weights $w(e)$ of the cycles passing through node $n+1$ through all edges $e=(c,n+1)$ for $c\leq n$ by $w(e)=P_{n,k}M_{c,l+1}M_{n+1,c}$.
	\item For each one of those edges, if the weight $w(e)$ is negative, flip the sign of $e$ with probability $p$.
	\item Increase $n$ by 1. If $n<N$, go back to step 2.
\end{enumerate}
The paths can be computed by $P_{n,k} = M_{n\times n} P_{n,k-1} - \text{diag}\left[M_{n\times n} P_{n,k-1}\right]$ where $\text{diag}\left[\cdot\right]$ is the operator that sets all non-diagonal entries to zero and the recursion is started by $P_{n,1}= M_{n\times n}$.
Combining correlations of different orders can be achieved by adding two matrices generated by this algorithm and normalizing.
The computational complexity of this algorithm is limited by the matrix multiplications needed to calculate the path weights, which has a complexity of $O(n^{2.373})$ and has to be called $n$ times at each of the $N$ iterations. This leaves us with $O(N^{4.373})$.

\section{Appendix B: Cavity method calculation for spectral density of regular cyclic graphs}

We derive the hypotrochoidic law for digraphs making use of an adaption of the cavity method \cite{rogers2009cavity,rogers2010spectral}, as used in \cite{metz2011spectra,bolle2013spectra}, coupled with an effective medium approximation (EMA) \cite{dorogovtsev2003spectra}. We start by expanding the Green's function
$G=(\mathcal{Z}-\mathcal{M})^{-1}$
around an arbitrary node $n$. Note that the matrix $\mathcal{M}$ describes the original graph $M$ and a replica with inverted connections $M^\dagger$.
Consider then the $2\times 2$ block of $G$ corresponding a node $n$ and the corresponding node $\tilde n=n+N$ in the replica graph. Using the Schur complement we can write
\begin{equation}
G_{n\tilde n, n\tilde n}=(Z-\mathcal{M}_{n\tilde n, \star}G^{(n)} \mathcal{M}_{\star, n \tilde n})^{-1}\,,
\end{equation}
where $\mathcal{M}_{n\tilde n, \star}$ refers to rows $n$ and $\tilde n$ of $\mathcal{M}$, $\mathcal{M}_{\star, n\tilde n}$ to the corresponding columns, and $G^{(n)}$ is the Green's function of the graph where nodes $n$ and  $\tilde n$ are removed. 

% Figure \ref{fig:nodefig}(a) illustrates the local neighbourhood of a node in the large graph limit with $k=3,d=3$. Each node is connected to $d$ pairs of nodes, shown in blue in the figure. With node $n$ removed, we must expand around a pair $p$ of nodes. As illustrated in Figure \ref{fig:nodefig}(b), in the absence of one neighbouring node (or pair) each pair is attached to $2(d-1)$ others, in two possible orientations.
% 
% \begin{figure}
% \hspace{-3cm}(a)\hspace{4cm}(b)\vspace{-5mm}
% \includegraphics[width=\columnwidth]{fig3.pdf}
% \caption{(a) Neighbourhood of a node in an oriented regular graph composed of 3-cycles, in which every node has in-degree and out-degree 3. (b) Neighbourhood of a pair of nodes (obtained by removing one node from a 3-cycle).}
% \label{fig:nodefig}
% \end{figure}

For graphs composed of $k$-cycles, where $k\geq3$, we develop our theory around pairing the source and drain endpoints of the segment $s$ of $k-1$ points that is obtained when removing the point $n$ from a single cycle. It is convenient to use indices $2$ to $k$ for these points. For the replica graph, $\tilde p$ represents the points $\tilde 2$ (now the drain) and $\tilde k$ (now the source). We are interested in the case where the node degrees and the cycle length $k$ are small relative to the graph size $N$, and nodes are assigned to cycles at random, such that the local topology approaches a tree-like digraph.

The Green's function of the graph $M$ on nodes $n$ and $\tilde n$ can then be written as
\begin{equation}
G_{n\tilde n,n \tilde n}=\bigg(Z-\sum_{m\sim n}G^{(n,m)}_{k\tilde 2,2\tilde k}\bigg)^{-1},
\end{equation}
where $G^{(n,m)}=(\mathcal{Z}-\mathcal{N})^{-1}$ is the Green's function of the graph with $n$ removed expanded around .

Truncated to the segments $s$ and $\tilde s$, this Green's function can be written as
\begin{equation}
G^{(n)}_{s\tilde s,s\tilde s}=
\Bigg[Z\otimes \openone_{k-1}-\mathcal{S}-(d-1)G^{(n)}_{k\tilde 2,2\tilde k}\otimes\openone_{k-1}\Bigg]^{-1}\,,
\end{equation}
where $\mathcal{S}=\mathrm{diag}\,(S,S^T)$ with
%\begin{equation}
$S_{i,j}=\delta_{i,j+1}$
%\end{equation}
represents the graph on the isolated segment $s$.
This gives the condition
\begin{equation}
\label{eq:gcond}
G^{(n)}_{s\tilde s,s\tilde s}=\left(
\begin{array}{cc}
\alpha-S & \beta \\
\gamma & \alpha^*-S^T \\
\end{array}
\right)^{-1}
\end{equation}
where, utilizing that $G^{(n)}_{\tilde 2, \tilde k}={G^{(n)}_{k,2}}^*$, we find
\begin{align}
\alpha&=z-(d-1)G^{(n)}_{k,2}, \label{eq:alpha} \\
\beta&=i\lambda-(d-1)G^{(n)}_{k,\tilde k}, \quad\gamma&=i\lambda-(d-1)G^{(n)}_{\tilde 2,2}\,.
\end{align}

On the boundary of the support of the spectrum, the off-diagonal elements $G^{(n)}_{k,\tilde k}$ and $G^{(n)}_{\tilde 2,2}$ vanish, while we always need to consider the limit $\lambda\to 0$. This means that $\beta$ and $\gamma$ become small, giving the expansion
\begin{equation}
G^{(n)}_{s\tilde s,s\tilde s}=\left(
\begin{array}{cc}
A & -\beta AA^\dagger \\
-\gamma A^\dagger A& A^\dagger \\
\end{array}
\right) +\mathcal{O}(\beta^2,\beta\gamma,\gamma^2),
\label{GnA}
\end{equation}
where the matrix $A$ has elements $A_{nm}=\alpha^{-1-n+m}$ if $n\geq m$ and 0 otherwise.
Comparing the matrix element $k,\tilde k$ on both sides gives the condition
\begin{equation}
G^{(n)}_{k,\tilde k}=(d-1)G^{(n)}_{k,\tilde k}\sum_{\ell=1}^{k-1}|\alpha|^{-2\ell},
\end{equation}
so that we can write $\alpha=t^{-1} e^{-i\varphi}$ where $t$ is the unique positive real solution of $(d-1)t^{2k}-dt^2+1=0$. Comparing the matrix element $k,2$ on both sides of (\ref{GnA}) we furthermore have the condition $G^{(n)}_{k,2}=\alpha^{1-k}$. Given the definition of $\alpha$ in Eq.~\eqref{eq:alpha}, we can express
$z=\alpha+(d-1)G^{(n)}_{k,2}$, so that
the boundary of the support is parameterized as
\begin{equation}
z_b(\varphi)=\frac{1}{t}e^{-i\varphi}+(d-1)t^{k-1}e^{i(k-1)\varphi}.
\label{eq:zb2}
\end{equation}
This again describes a hypotrochoid curve, and is in agreement with numerical results as presented in Fig.~\ref{hyps}.

\section{Appendix C: Derivation for graphs with multiple cycle structures}
For a graph where each vertex connects to $d_1$ $k_1$-cycles of weight $w_1$ and
$d_2$ $k_2$-cycles of weight $w_2$, the Green's functions on a node $n$ and its replica $\tilde n$ can be written as
\begin{equation}
G_{n\tilde n,n \tilde n}=(Z-d_1w_1^2 G^{(1)}_{k_1\tilde 2,2\tilde k_1}-d_2w_2^2 G^{(2)}_{k_2\tilde 2,2\tilde k_2})^{-1}
\end{equation}
with
\begin{widetext}
\begin{align}
G^{(1)}_{s_1\tilde s_1,s_1\tilde s_1}=
\Bigg[Z\otimes \openone_{k_1-1}-\mathcal{S}^{(1)}-(d_1-1)w_1^2G^{(1)}_{k_1\tilde 2,2\tilde k_1}\otimes\openone_{k_1-1}-d_2w_2^2G^{(2)}_{k_2\tilde 2,2\tilde k_2}\otimes\openone_{k_1-1}\Bigg]^{-1},
\\
G^{(2)}_{s_2\tilde s_2,s_2\tilde s_2}=
\Bigg[Z\otimes \openone_{k_2-1}-\mathcal{S}^{(2)}-d_1 w_1^2G^{(1)}_{k_1\tilde 2,2\tilde k_1}\otimes\openone_{k_2-1}-(d_2-1)w_2^2G^{(2)}_{k_2\tilde 2,2\tilde k_2}\otimes\openone_{k_2-1}\Bigg]^{-1},
\end{align}
\end{widetext}
where the isolated segments $s_i$ with adjacency matrix $S^{(i)}$ now occupy points $2$ to $k_i$.

These matrices are still of the form \eqref{eq:gcond}, where now
\begin{subequations}\label{eqs:abgdef}
	\begin{align}
	\alpha_1=z-(d_1-1)w_1^2G^{(1)}_{k_1,2}-d_2w_2^2G^{(2)}_{k_2,2}, \\
	\alpha_2=z-d_1 w_1^2G^{(1)}_{k_1,2}-(d_2-1)w_2^2G^{(2)}_{k_2,2}, \\
	\beta_1=i\lambda-(d_1-1)w_1^2G^{(1)}_{k_1,\tilde k_1}-d_2w_2^2G^{(2)}_{k_2,\tilde k_2}, \\
	\beta_2=i\lambda-d_1w_1^2G^{(1)}_{k_1,\tilde k_1}-(d_2-1)w_2^2G^{(2)}_{k_2,\tilde k_2}, \\
	\gamma_1=i\lambda-(d_1-1)w_1^2G^{(1)}_{\tilde 2,2}-d_2w_2^2G^{(2)}_{\tilde 2,2}, \\
	\gamma_2=i\lambda-d_1w_1^2G^{(1)}_{\tilde 2,2}-(d_2-1)w_2^2G^{(2)}_{\tilde 2,2}.
	\end{align}
\end{subequations}

On the boundary it now is consistent to require that both $\beta_1$ and $\beta_2$ are small. Inverting the matrices perturbatively as before, we then find the following conditions,
\begin{subequations}\label{eqs:gcond2}
	\begin{align}
	G^{(1)}_{k_1,2}&=w_1^{k_1-2}\alpha_1^{1-k_1},\\
	G^{(2)}_{k_2,2}&=w_2^{k_2-2}\alpha_2^{1-k_2},\\
	G^{(1)}_{k_1,\tilde k_1}&=[(d_1-1)w_1^2G^{(1)}_{k_1,\tilde k_1}+d_2w_2^2G^{(2)}_{k_2,\tilde k_2}]w_1^{-2}\Sigma_1,\\
	G^{(2)}_{k_2,\tilde k_2}&=[d_1w_1^2G^{(1)}_{k_1,\tilde k_1}+(d_2-1)w_2^2G^{(2)}_{k_2,\tilde k_2}]w_2^{-2}\Sigma_2,
	\end{align}
\end{subequations}
where  the last two require
\begin{equation}
(-d_1-d_2+1)\Sigma_1\Sigma_2-(d_1-1)\Sigma_1-(d_2-1)\Sigma_2+1=0,
\label{eq:condt}
\end{equation}
with
$\Sigma_r=\sum_{l=1}^{k_r-1}|\alpha_r/w_r|^{-2l}$.
Furthermore, using the definitions of $\alpha_1$ and $\alpha_2$ in Eq.~\eqref{eqs:abgdef} and the first two conditions in Eq.~\eqref{eqs:gcond2}
we obtain
\begin{align}
z&=\alpha_1+(d_1-1)w_1^{k_1}\alpha_1^{1-k_1}+d_2w_2^{k_2}\alpha_2^{1-k_2}\\
&=\alpha_2+d_1w_1^{k_1}\alpha_1^{1-k_1}+(d_2-1)w_2^{k_2}\alpha_2^{1-k_2},
\end{align}
which have to agree,
hence
\begin{equation}
\alpha_1-\alpha_2=w_1^{k_1}\alpha_1^{1-k_1}-w_2^{k_2}\alpha_2^{1-k_2}.
\label{eq:extracond}
\end{equation}
We now set $\alpha_r=(w_r/t_r)\exp(-i\varphi_r)$, so that
\begin{align}
\Sigma_1&=\sum_{l=1}^{k_1-1}t_1^{2l},
\quad
\Sigma_2=\sum_{l=1}^{k_2-1}t_2^{2l},
\end{align}
Equation \eqref{eq:condt} then relates $t_2$ to $t_1$,
while Eq.~\eqref{eq:extracond} further relates $\varphi_2$ to $\varphi_1$.
Given these relations, the boundary curve can then be written, in symmetrised form, as
\begin{widetext}
\begin{align}
z&=\frac{w_1}{2t_1}\exp(-i\varphi_1)+\frac{w_2}{2t_2}\exp(-i\varphi_2)+(d_1-1/2)w_1t_1^{k_1-1}\exp(i(k_1-1)\varphi_1)
+(d_2-1/2)w_2t_2^{k_2-1}\exp(i(k_2-1)\varphi_2).
\end{align}
\end{widetext}

Explicit expressions are obtained for $d_1,d_2\gg 1$. Then, conditions \eqref{eq:condt} and  \eqref{eq:extracond} give in leading order $\varphi_2=\varphi_1\equiv\varphi$ and
\begin{align}
t_1&\sim \frac{w_1}{\sqrt{d_1w_1^2+d_2w_2^2}}, \quad t_2\sim \frac{w_2}{\sqrt{d_1w_1^2+d_2w_2^2}},
\end{align}
so that
\begin{align}
\frac{z}{\bar d}&=e^{-i\varphi}+d_1\left(\frac{w_1}{\bar d}\right)^{k_1}e^{i(k_1-1)\varphi}
+d_2\left(\frac{w_2}{\bar d}\right)^{k_2}e^{i(k_2-1)\varphi},
\end{align}
where $\bar d=\sqrt{d_1w_1^2+d_2w_2^2}$.

\bibliography{spectraCyclesBibliography}

\end{document}